# A Reversible Data hiding Scheme in Encrypted Domain for Secret Image Sharing based on Chinese Remainder Theorem

Yan Ke, Minqing Zhang, Xinpeng Zhang, *Member*, *IEEE*, Jia Liu, Tingting Su, and Xiaoyuan Yang

*Abstract*—Reversible data hiding in encrypted domain (RDH-ED) schemes based on symmetric or public key encryption are mainly applied to the security of end-to-end communication. Aimed at providing reliable technical supports for multi-party security scenarios, a separable RDH-ED scheme for secret image sharing based on Chinese remainder theorem (CRT) is presented. In the application of (*t*, *n*) secret image sharing, the image is first shared into *n* different shares of ciphertext. Only when not less than *t* shares obtained, can the original image be reconstructed. In our scheme, additional data could be embedded into the image shares. To realize data extraction from the image shares and the reconstructed image separably, two data hiding methods are proposed: one is homomorphic difference expansion in encrypted domain (HDE-ED) that supports data extraction from the reconstructed image by utilizing the addition homomorphism of CRT secret sharing; the other is difference expansion in image shares (DE-IS) that supports the data extraction from the marked shares before image reconstruction. Experimental results demonstrate that the proposed scheme could not only maintain the security and the threshold function of secret sharing system, but also obtain a better reversibility and efficiency compared with most existing RDH-ED algorithms. The maximum embedding rate of HDE-ED could reach 0.5000 bits per pixel and the average embedding rate of DE-IS is 0.0545 bits per bit of ciphertext.

## I. Introduction

REVERSIBLE data hiding in encrypted domain (RDH-ED) is an information hiding technique that can not only embed and extract additional data in the ciphertext, but also restore the plaintext losslessly [1][2]. RDH-ED is useful in distortion-unacceptable scenarios, such as cloud data management or ciphertext management for medical or military use [1]. With the increasing demands for information security, various applications of encryptions continue to be popularized, leading to an increasing share of encrypted data in cyberspace, which puts forward more urgent demands for practicable and versatile RDH-ED techniques.

Existing cryptosystems that have been introduced into RDH-ED mainly includes three categories: *Symmetric encryption* [3]-[6], *public key encryption* [1], and *secret sharing* [7]. Symmetric or public key encryption is mainly used to ensure the security of end-to-end communication. However in nowadays, *multi-party security* has been occupying an increasingly significant position, such as in secure multi-party computing (MPC), block chain, and federated learning [8]. As an important cryptosystem for multi-party security, secret sharing has been widely applied to secure MPC, key agreement, digital signature, transfer and voting systems [8] [9].

The core function of the secret sharing system lies in *the (t, n) threshold* of it ($n > t \geqslant 2$) [8]. A (*t*, *n*)-threshold secret sharing scheme is a fundamental *cryptographic primitive*, which allows a *dealer* owning a secret to distribute this secret among a group of *n shareholders* in such a way that any *t* shareholders can reconstruct the secret, but no subset of less than *t* shareholders can gain any information on the secret. In this way, *access control* to a system can be enforced, because the secret can be maintained secrecy even against an attacker who manages to break into/eavesdrop up to less than *t* shareholders of the system. Another advantage of (*t*, *n*) secret-sharing schemes is to achieve **robustness** of secret reconstruction [10], since the secret can be recovered as long as there are *t* shares available.

Secret image sharing (SIS) is an important part of research in multi-party security cryptosystems [10][11], which could realize access control and share the risk of single point failure of the system to improve the robustness of secret image reconstruction. Processes of (3, 5) SIS are demonstrated briefly in Fig. 1. The secret image is owned by the dealer of a system and shared into 5 image shares. The image shares are meaningless and then distributed to 5 independent servers as shown in Fig. 1(a). Only when any 3 of the shares obtained, can the secret image be reconstructed as shown in Fig. 1(b), thus ensuring the *access control* and *robustness* of image reconstruction. In current researches, multi-party systems not only require that the dealer could share and reconstruct the secret image robustly, but also demand the identity authentication, integrity checking, and other managements in SIS, as stressed in [12]-[15], which provides promising application prospects for RDH-ED.

Application instances of RDH-ED for SIS are shown in Fig.

This work was supported in part by National Natural Science Foundation of China under Grant 61872384 and Grant U1603261, and National Key R&D Program of China under Grant 2017YFB0802000.

Yan Ke is with Counterterrorism Command & Information Engineering Joint Lab in Urumqi campus of Engineering University of PAP, Urumqi, 830049, China and the Key Laboratory of Network and Information Security under PAP in Engineering University of PAP, Xi'an, 710086, China (e-mail: 15114873390@163.com).

Minqing Zhang(*Corresponding author*), Jia Liu, Tingting Su, and Xiaoyuan Yang are with the School of Cryptography Engineering in Engineering University of PAP, Xi'an, 710086, China (e-mail: api_zmq@126.com; liujia1022@gmail.com; suting 0518@163.com; yxyangyxyang@163.com).

Xinpeng Zhang is with School of Computer Science, Fudan University, Shanghai, 200000, China (e-mail: zhangxinpeng@fudan.com).

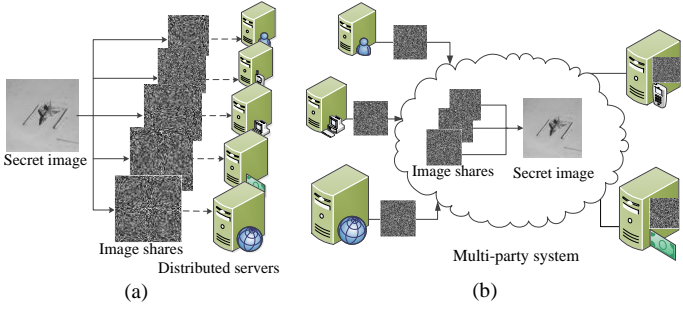

Fig. 1. An overview of (3, 5)-threshold secret image sharing: (a) Secret image sharing and distribution; (b) Secret image reconstruction.

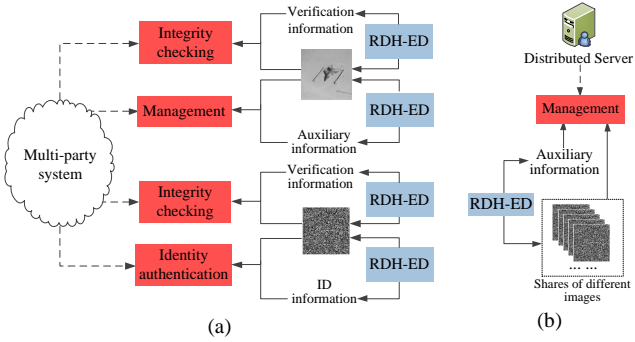

Fig. 2. Application instances of RDH-ED for SIS. (a) For the dealer of the system: integrity checking, identity authentication, and management on shares and secret images; (b) For the distributed servers: management on shares.

2. Additional data applied to various functions (e.g., integrity checking, identity authentication, or management, etc.) can be embedded into the shares and the reconstructed secret images by RDH-ED. For the multi-party system as shown in Fig. 2(a), the dealer of the system could authenticate the identity and check the integrity of the shares uploaded from different servers by using the embedded ID information and verification information. The dealer could also check the integrity of the reconstructed image by using the embedded verification information, and manage the images by using the embedded auxiliary information. For each server, it could manage the shares from different secret images by using the embedded auxiliary information as shown in Fig. 2(b).

In the proposed application of RDH-ED for SIS, the embedding operations are implemented on the image shares to ensure the confidentiality of the image content to servers. After embedding, the server stores the marked share instead of the original one locally, which cannot reveal any information about the secret image. To achieve flexibility and practicality of the managements for SIS, the data extraction processes should be separable, that is, *a)* the server and the dealer of the system could extract data directly from the marked shares; *b)* the dealer of the system could extract data from the secret image reconstructed from the marked shares, namely, the directly reconstructed image carries additional data. Therefore, there is distortion in the reconstructed image, and a restoration process is needed to restore the original secret image losslessly.

## II. RELATED WORK

With more cryptosystems being introduced, the integration and reference of data hiding techniques and cryptography techniques have been more and more profound.

Typical *symmetric encryptions* in RDH-ED are stream encryption [3]-[6], advanced encryption standard (AES) [16][17], and Rivest cipher 4 (RC4) encryption [18].

The first stream encryption based RDH-ED method was proposed by Zhang for encrypted images [3]. Scheme in [5] enhanced the embedding capacity (EC) based on [3]. Qian *et al.* proposed a method to realizing data hiding in the encrypted JPEG bit streams [6]. AES was introduced by Puech *et al.* [16] to encrypt the image before embedding. And then difference prediction was used as preprocessing before encryption in [17], thus enhancing the EC and reversibility of [16]. However, all above RDH-ED methods are not separable, namely, the order of data extraction and decryption operations are fixed, which restricted the practicability of RDH-ED. Separable RDH-ED was first proposed in [19]. Separability has been so far an important attribute of practicality for current RDH-ED [1][20].

The difficulty of realizing reversible data hiding in the symmetric encryption domain lies in that traditional embedding redundancy from the correlation of plaintext would be destroyed in the ciphertext. To introducing redundancy into ciphertext, methods of "vacating room before encryption (VRBE)" [3], [5]-[16]and "vacating room after encryption (VRAE)" [4][17] were proposed. However, the embedding redundancy introduced by VRAE or VRBE is independent from the encryption, resulting in mutual restriction between decryption distortion and EC. To better utilize the redundancy, the first solution is to improve the correlation of redundancy in VRBE. Puech in [21] proposed a high EC algorithm with by making full use of prediction error of the most significant bits (MSB) before encryption. The other solution is to preserve correlation of the plaintext, so that RDH for spatial domain, e.g., difference expansion technique (DE) [23], histogram shifting technique (HS) [24]-[25], could be implemented in encrypted domain. Huang in [22] proposed a VRAE framework of RDH-ED, in which the same random sequences were reused as the encryption keys to preserve the correlation among neighboring pixels in the ciphertext of pixels. The RC4 encryption was declared breached in 2013 [26], RDH-ED based on early RC4 has limitations in security applications.

Symmetric encryption based RDH-ED algorithms are fast and efficient, the technical goals of which are to further increase the EC and to realize separability and the security of RDH-ED [1]. The defect of symmetric encryption in practice is the cost of key's distribution and storage is high for each user. As a comparison, the key storage cost of public key encryption is relatively low. Public key encryption based RDH-ED has been so far an emerging direction of the research [1][2].

*Public key encryption* based RDH-ED are mainly based on Paillier encryption [27]-[33] and learning with Error (LWE) encryption [34]-[36].

The first Paillier encryption based RDH-ED was proposed by Chen *et al.* in [27], which is VRBE scheme. Shiu *et al.* [28] and Wu *et al.* [29] improved the EC in [27] by solving the pixel overflow problem. Li *et al.* in [30] proposed a VRAE scheme with a considerable EC by utilizing the homomorphic addition property of Paillier encryption and HS technique. All above schemes were all inseparable. Data extraction was only in the plaintext domain. The separable VRAE schemes were proposed in [31], [32]. Wu *et al.* proposed two data hiding algorithms for encrypted images in [31]. One was based on Paillier cryptosystem for data extraction after decryption, and the other supported data extraction in the encryption domain. Zhang *et al.* [32] proposed a combined scheme consisting of a lossless scheme and a reversible scheme to realize separability. Xiang proposed a separable VRBE scheme in [33]. Data was embedded into the LSBs of pixels by employing homomorphic multiplication. The one-to-one mapping table from ciphertext to plaintext was introduced for data extraction from the encrypted domain.

Learning with errors (LWE) encryption was first introduced by Zhang and Ke in [34] by quantifying the LWE encrypted domain and recoding the LWE ciphertext. Ke *et al.* fixed the parameters for LWE encryption and proposed a multilevel RDH-ED with a higher EC in [35]. In [36], correlation from the plaintext was preserved in the ciphertext through a modified somewhat-LWE encryption for separability. Ke in [1] proposed a secure separable scheme by using the fully homomorphic encryption and key-switching technique. All in all, the public key encryption based RDH-ED concentrates on the implementation of separability, high EC and security of RDH-ED.

Symmetric or public key encryptions are mainly for end-to-end communication systems. Besides that, RDH-ED could also provide reliable technical supports of data hiding for multi-party applications of secret sharing as discussed in Section I.

*Secret sharing* was first introduced into RDH-ED by Wu *et al.* in [7] by using Shamir secret sharing to share an image. Additional data was embedded into the image shares by DE and HS techniques. Scheme [7] is inseparable and the operation of image reconstruction must follow data extraction. Chen *et al*, [9] used a (2, 2) SIS scheme to share image pixels and embedded additional data into the shares by utilizing homomorphic addition. The (2, 2) SIS scheme in [9] cannot retain the threshold function of secret sharing. Data can only be extracted after image reconstruction.

This paper proposes a separable RDH-ED scheme for secret image sharing based on Chinese remainder theorem (CRT), which includes two data hiding processes to support data extraction before and after image reconstruction. The proposed scheme maintains ($t$, $n$) threshold of secret image sharing and obtains a better EC compared with the existing RDH-ED algorithms.

The rest of this paper is organized as follows. The following section introduces the processes of CRT secret sharing. Section IV introduces the proposed RDH-ED scheme for image sharing. In Section V, the three judging standards of RDH-ED, including *correctness*, *security* and *efficiency*, are discussed theoretically and verified with experimental results. Finally, Section VI summarizes the paper and discusses future investigations.

## III. SECRET SHARING SCHEME BASED ON CRT

The standard integer-based Chinese remainder theorem ($t$, $n$) secret sharing scheme was defined by Asmuth and Bloom in 1980 [37], [38]:

### A. Parameters Generation

Pick a (not necessarily random) prime $q_0$ as module for secret reconstruction. Generate $n$ distinct random primes $\boldsymbol{q} = (q_1,..., q_n)$ as modules for secret sharing in ascending order, where $q_i < q_j$, if $i < j$. All the modules meet the following conditions:

*a)* $m$ is the secret to be shared:
$$q_0 > m \tag{1}$$
*b)* Any two modules are with coprime:
$$\gcd(q_i, q_j)=1, \ \forall \ i, j \in [0, n], i \neq j. \tag{2}$$
*c)* $u = \prod_{i=1}^{t} q_i > q_0 \prod_{j=1}^{t-1} q_{n-j+1} \tag{3}$

### B. Secret Sharing

Let $u = \prod_{i=1}^{t} q_i$, to share secret $m \in Z_{q_0}$, choose a uniformly random $r \in [0, u)$. Compute the integer $g = m + r \cdot q_0$. The *i*-th share $(q_i, s_i)$ would be transported to the *i*-th user: $s_i = g \mod q_i$ for $i = 1, 2, ..., n$.

### C. Secret Reconstruction

At least $t$ shares from the $n$ users at random are required to reconstruct the secret. The $t$ shares are denoted as: $\{(q_i', s_i')| i=1, 2,…,t\}$.

The unique $g'$ is computed by Chinese remaindering [37] to meet $s_i'= g' \mod q_i'$ for $i =1, 2,…, t$. The reconstructed secret is obtained:
$$s = g' \mod q_0 \tag{4}$$

### D. Homomorphic Addition

Currently, the typical homomorphic algorithms of public key encryption are Paillier algorithm with multiplication homomorphism, ECC (Elliptic Curve Cryptography) with addition homomorphism, and LWE encryption with full homomorphism, etc. Moreover, several symmetrical encryptions have also addition homomorphism, such as RC4 or stream cipher, based on which RDH-ED methods [3][18] were proposed. The common basis of homomorphic symmetric encryption is that the encryptions are based on linear addition and modulus operations. For example, the XOR in stream cipher is the operation of addition modulo 2. The addition homomorphism of CRT secret sharing is also based on the operations of addition modulo $q$, which can be described as following:

Let $a = b \mod q$, $c = d \mod q$, then $a + c = (b + d) \mod q$.

For two secrets $m, m'$ and their shares $\{s_1, …, s_n\}$, $\{s_1', …, s_n'\}$ by CRT secret sharing, the addition homomorphism ensrues that the reconstructed secret of share $\{s_1 + s_1', …, s_n + s_n'\}$ is $m + m'$。

Two classical constructions for ($t$, $n$) secret sharing are CRT secret sharing scheme [37][38] and the Shamir polynomial based secret sharing scheme [39]. In this paper, we focus on the realization of RDH-ED based on CRT. Homomorphic difference expansion in encrypted domain of CRT secret sharing is proposed by utilizing the homomorphic addition of CRT, which could also be applied in Shamir schemes in our future work.

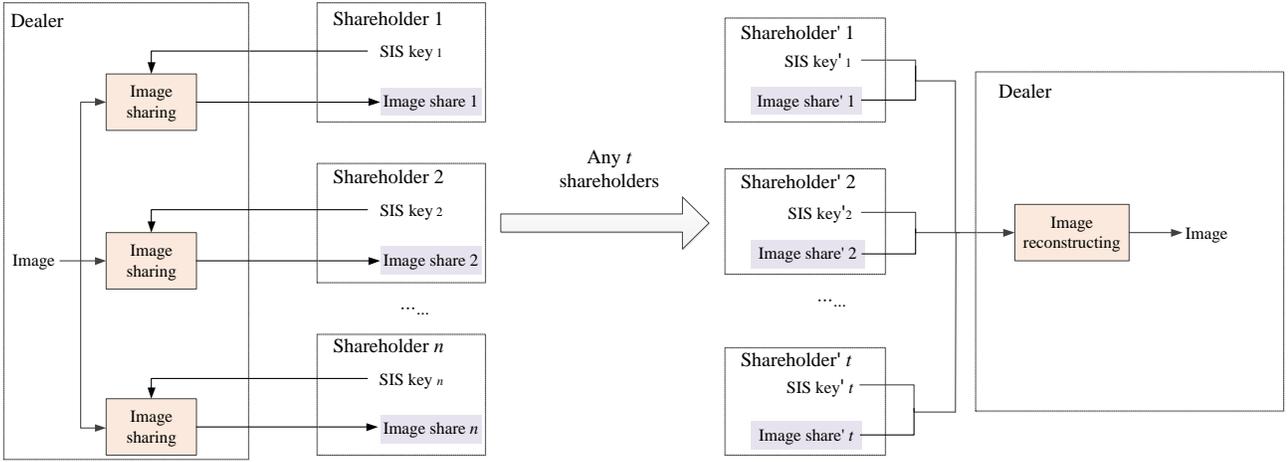

Fig. 3. The processes of secret image sharing.

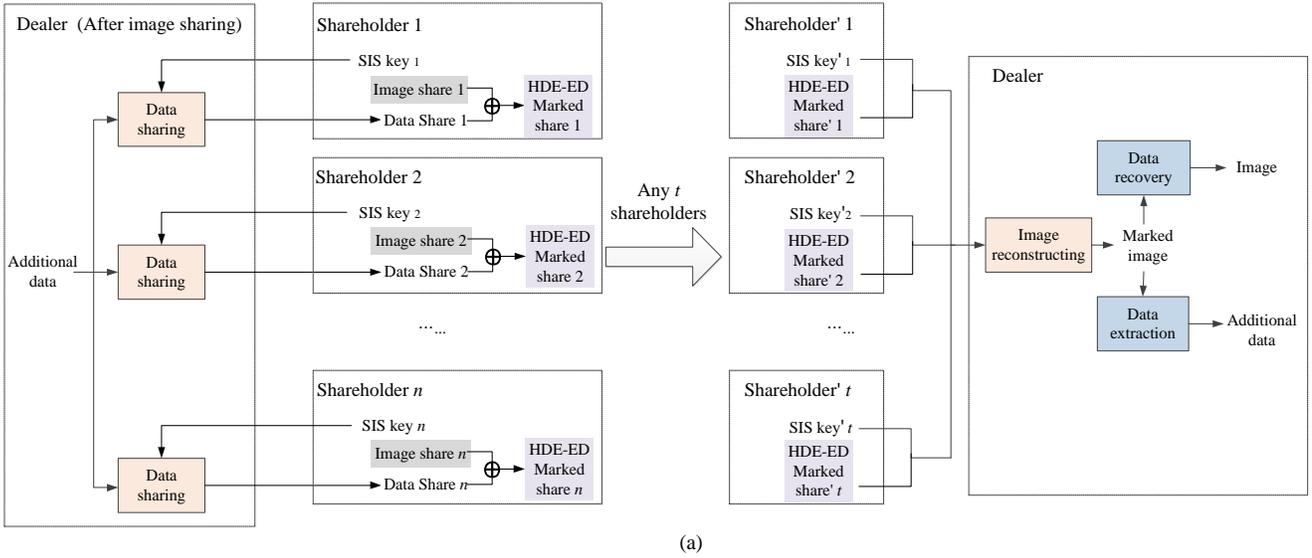

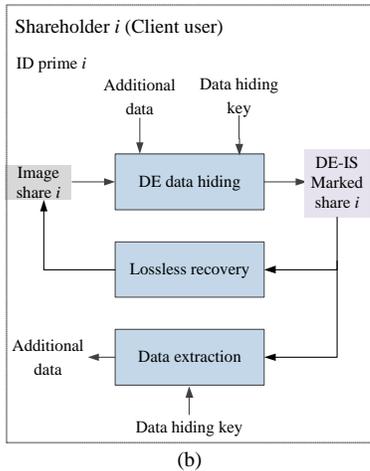

(b)

Fig. 4. Frameworks of the proposed two RDH-ED methods for SIS: (a) Homomorphic difference expansion in encrypted domain (HDE-ED); (b) Difference expansion in image shares (DE-IS).

## IV. THE PROPOSED SCHEME

### A. Framework and Notation

In the application of secret image sharing, there are $n$ different shareholders and a dealer of the system. The dealer owns the content of the secret image and is supposed to set SIS system parameters, generate and distribute primes as the SIS key to each shareholder. The dealer shares the image and sends the shares to shareholders for distributed storage. Combined with Fig. 1-3, the dealer in Fig. 3 corresponds to the multi-party system in Fig. 1-2 while the shareholders correspond to the distributed servers.

The processes of SIS are as shown in Fig. 3. The shareholder $i$ is distributed the SIS key $i$ in advance ($i$=1, 2,..., $n$).To calculate the image share $i$, the dealer must receive the shareholder $i$' SIS key, and input the key together with the image into image sharing function. Each shareholder keeps his image share and SIS key locally. To reconstruct the image, at least $t$ shareholders must provide their shares and SIS keys to the dealer to restore the image.

Based on SIS, we propose two RDH-ED methods to realize reversible embedding in the image shares. The separability is embodied in that the additional data can be extracted not only from the marked shares by the shareholders before reconstruction, but also from the reconstructed image by the dealer. The frameworks of the proposed methods are shown in Fig. 4:

The first is *homomorphic difference expansion in encrypted domain* (HDE-ED) as shown in Fig. 4(a). The additional data is shared into $n$ data shares. HDE-ED marked share could be obtained by operating homomorphic addition between the image share and data share. HDE-ED ensures the dealer could obtain a marked image with additional data embedded after image reconstruction. Additional data could be extracted from the marked image, and the original image can be restored. The proposed HDE-ED is operated in the encrypted domain of image shares, which could realizes homomorphic embedding modifications on the image content without reconstructing the image.

The second is *difference expansion in image shares* (DE-IS) as shown in Fig. 4(b). DE-IS ensures that any a shareholder could embed additional data into his image share by expanding the differences between the ciphertext of image share and his SIS key. Additional data can be extracted from DE-IS marked share by using data-hiding key, and the image share could be losslessly recovered. DE-IS is also implemented on the image share, which reveals nothing about the secret image. The recovered image share is the same as the image share before DE-IS embedding.

The secret image can be of any size. In this paper, we use 512×512 8-bit grayscale image as the secret image to facilitate the introduction of the proposed scheme. The notation of the main variables in the scheme is shown in Table I.

### B. Parameter Setting and Key Distribution

$q_0$ is the modulus used for image reconstruction, thus resulting in a recoverable pixel value range of $[0, q_0-1]$. For 8-bit grayscale images with a pixel range of $[0, 255]$, there would be pixels unrecovered in the reconstructed image if $q_0$ takes a value less than 255 according Eq. (1). On the other hand, if $q_0$ takes a value greater than 255, there would be data expansion in the ciphertext of image shares according to Eq. (3), in which $q_i$ ($i=1, 2,…, n$) > $q_0$ and $q_i$ is the modulus used for image sharing. $q_i$ determines the range of ciphertext in the image share. Traditional image sharing schemes choose $q_0<255$ to ensure no data expansion brought in. They do not require all pixels restored in the reconstructed image due to the redundancy in the correlation of image content. To restore the image losslessly and reduce the data extension of ciphertext as much as possible, the parameter settings of the proposed RDH-ED scheme are as follows:

Pick a prime $q_0 \in [2^w, 2^{w+1}]$ as the reconstruction module and more than $n$ modules: $\boldsymbol{q} = (q_1, q_2, …) \in [2^w, 2^{w+1}]$. In this paper, $w$ is set 8 since $\boldsymbol{I}$ is an 8-bit grayscale image. To share a pixel, $n$ modules from $\boldsymbol{q}$ would be randomly selected and distributed to $n$ shareholders as the prime for image sharing. For security, one prime should only be used by one shareholder for sharing one pixel at a time.

Therefore, each shareholder needs to be distributed a prime matrix of size 512×512 in advance as the SIS key. The SIS key of the $i$-th shareholder is denoted as $\boldsymbol{K}_{SIS}^i$, and the prime is denoted as $ID_{x,y}^i \in \boldsymbol{K}_{SIS}^i$, $i=1, 2,…, n$; $x, y \in \{1, …, 512\}$. Choose a 512×512 random matrix $\boldsymbol{R} \in Z_{\lfloor u/q_0 \rfloor}^{512 \times 512}$, $r_{x,y} \in \boldsymbol{R}$, $x, y \in \{1, …, 512\}$. The public parameters are $\boldsymbol{R}$ and $q_0$. Parameters and keys are distributed as shown in Table II.

### C. Homomorphic Difference Expansion in Encrypted Domain

The purpose of HDE-ED is to ensure that *a)* the reconstructed image would carry additional data through certain homomorphic processing on the image shares of the $n$ shareholders. *b)* The original image could be recovered from the reconstructed marked image.

#### 1) Constraints for Difference Expansion

The to-be-shared image is a 512×512 image $\boldsymbol{I}$. $\boldsymbol{I}$ is divided into non-overlapping pixel pairs. We take the pixel pair as the basic unit to scramble $\boldsymbol{I}$, aiming to destroy the structural correlation of it.

Each pair consists of two adjacent pixels $(P_{x,y}, P_{x,y+1})$, $x \in \{1, 2, 3,…, 512\}$, $y \in \{1, 3, 5,…, 511\}$. Assuming $P_{x,y} > P_{x,y+1}$, the difference $h_{x,y}$ and average value $l_{x,y}$ (integer) of $P_{x,y}$ and $P_{x,y+1}$ are computed as following:

$$h_{x,y} = P_{x,y} - P_{x,y+1} \qquad (5)$$

$$l_{x,y} = \left\lfloor \frac{P_{x,y} + P_{x,y+1}}{2} \right\rfloor \qquad (6)$$

TABLE I
NOTATION

| Denotation | Representation |
|---|---|
| $\boldsymbol{I}$ | The 512×512 image for sharing. |
| $P_{x,y} \in Z_{256}$ | The pixel in row $x$, column $y$ of an image ($x, y \in \{1, …, 512\}$). |
| $S_i$ ($i=1, …, n$) | The $i$-th share of image $\boldsymbol{I}$. |
| $c_{x,y}^i \in Z_{qi}$ ($i=1, …, n$) | The ciphertext in row $x$, column $y$ of the share $S_i$, $x, y \in \{1, …, 512\}$. |
| $b_s \in Z_2$ | The additional data to be embedded. |
| $S_i'$ ($i=1, …, n$) | The marked share by HDE-ED embedding. |
| $S_i''$ ($i=1, …, n$) | The marked share by DE-IS embedding. |
| $c_{x,y}^i{'} \in Z_{qi}$ ($i=1, …, n$) | The marked ciphertext in row $x$, column $y$ of the share $S_i'$, $x, y \in \{1, …, 512\}$. |
| $c_{x,y}^i{''} \in Z_{qi}$ ($i=1, …, n$) | The marked ciphertext in row $x$, column $y$ of the share $S_i''$, $x, y \in \{1, …, 512\}$. |
| $\boldsymbol{I}'$ | The 512×512 marked image reconstructed from any $t$ marked image shares. |
| $P_{x,y}' \in Z_{256}$ | The pixel in row $x$, column $y$ of the marked image $\boldsymbol{I}'$, $x, y \in \{1, …, 512\}$. |

TABLE II
KEY DISTRIBUTION

| Classification | Denotation | Function | Owner |
|---|---|---|---|
| SIS key | $\boldsymbol{K}_{SIS}^i$ ($i=1,…,n$) | Image sharing & reconstruction | The $i$-th shareholder |
| Public parameter | $\boldsymbol{R}$, $q_0$ | Image sharing | Open to the public |
| Data-hiding key | $k$ | DE-IS | The $i$-th shareholder |

$$P_{x,y} = l_{x,y} + \left\lfloor \frac{h_{x,y}+1}{2} \right\rfloor \quad (7)$$

$$P_{x,y+1} = l_{x,y} - \left\lfloor \frac{h_{x,y}}{2} \right\rfloor \quad (8)$$

$\lfloor . \rfloor$ is the floor function meaning "the biggest integer less than or equal to". As grayscale values are bounded in [0, 255], we have constraints about $h_{x,y}$ and $l_{x,y}$ according to Eqs. (7), (8):

$$0 \leq l_{x,y} + \left\lfloor \frac{h_{x,y}+1}{2} \right\rfloor \leq 255 \quad (9)$$

$$0 \leq l_{x,y} - \left\lfloor \frac{h_{x,y}}{2} \right\rfloor \leq 255 \quad (10)$$

To avoid overflow or underflow problems after data hiding, The difference $h_{x,y}$ of an available pair should satisfy the following constraints [23]:

$$|h_{x,y}| \leq \min(2(255-l_{x,y}), 2l_{x,y}+1) \quad (11)$$
$$|2 h_{x,y}+b_s| \leq \min(2(255-l_{x,y}), 2l_{x,y}+1) \quad (12)$$

for $b = 0$ or 1.

We add an extra fidelity constraint: the pixel pairs are preferentially used for embedding with a smaller pixel difference. The fidelity parameter is denoted as $h_{fid}$ [1]:

$$h_{x,y} \leqslant h_{fid} \quad (13)$$

For all the pixel pair $(P_{x,y}, P_{x,y+1})$, the difference and average would replace the pixel value for the following processing:

$$P_{x,y} = h_{x,y} \quad (14)$$
$$P_{x,y+1} = l_{x,y} \quad (15)$$

Before image sharing, we use a map matrix $M_{ava} \in \{0,1\}^{512 \times 512}$ to label available pixel pairs of the scrambled $I$. Value "1" indicates the bigger pixel within an available pixel pair for DE embedding ($M_{ava}$ would be lossless compressed as side information of the ciphertext to superimpose on the host signal [1].)

*2) Image Sharing*

For the $i$-th shareholder ($i=1, 2,..., n$), the image share $S_i$ would be generated by using ($I$, $K_{SIS}^i$).

For an pixel $P_{x,y}$, $x, y \in \{1, 2, 3,..., 512\}$, compute the integer $g_{x,y} = P_{x,y} + r_{x,y} q_0$. Then $c_{x,y}^i$ in $S_i$ would be obtained:

$$c_{x,y}^i = g_{x,y} \bmod ID_{x,y}^i \quad (16)$$

*3) HDE-ED Marked Image Share Generation*

The additional bit $b_s$ would be embedded into the available pair $(P_{x,y}, P_{x,y+1})$, i.e., $(h_{x,y}, l_{x,y})$, according to Eqs. (14)-(15), $x \in \{1, 2, 3,..., 512\}$, $y \in \{1, 3, 5,..., 511\}$.

Step 1: *Additional data sharing*. For the $i$-th shareholder ($i=1, 2,..., n$), the data share is denoted as $d_{x,y}$. It would be generated by using ($b_s$, $ID_{x,y}^i$):

$$d_{x,y} = b_s \bmod ID_{x,y}^i \quad (17)$$

Step 2: *Marked image share generation*. For all the $n$ share of image $I$,

$$c_{x,y}^{i\prime} = (c_{x,y}^i + d_{x,y}) \bmod ID_{x,y}^i \quad (18)$$

where $i = 1, 2,..., n$.

For all the available pairs, each pair would carry one bit additional data by repeating the operations in Eqs. (17)-(18).

The $i$-th shareholder would obtain HDE-ED marked share $S_i'$ ($i=1, 2,..., n$).

*4) Marked Image Reconstruction*

With any $t$ shareholders and their shares $S_i'$ ($i=1, 2,..., t$), the marked image $I'$ could be obtained by reconstructing all the pixels of $I'$ by using the SIS key $K_{SIS}^i$.

As for $P_{x,y}'$, $x, y \in \{1, ..., 512\}$, $g''$ is first computed by Chinese Remaindering to meet:

$$c_{x,y}^{i\prime} = g'' \bmod ID_{x,y}^i, \quad i=1, 2,..., t \quad (19)$$

$$P_{x,y}' = g'' \bmod q_0 \quad (20)$$

Then the marked difference and average could be obtained:

$$h_{x,y}' = P_{x,y} \quad (21)$$
$$l_{x,y}' = P_{x,y+1} \quad (22)$$

where $x \in \{1, 2, 3,..., 512\}$, $y \in \{1, 3, 5,..., 511\}$.

The marked pixel pair could be recovered according to Eqs. (7)-(8). Then the marked image $I'$ could be obtained by inversely scrambling the pixel pairs.

*5) Data Extraction and Image Recovery*

With the marked images $I'$, the data extraction and image recovery follows the traditional DE method. As for the marked pixel pair ($P_{x,y}'$, $P_{x,y+1}'$), $h_{x,y}'$ is computed according to Eq. (5). $b_s$ could be extracted:

$$b_s = LSB(h_{x,y}') \quad (23)$$

$LSB(.)$ is to obtain the least significant bit of the input integer.

To recover the image, the original difference should be recovered first:

$$h_{x,y} = \lfloor h_{x,y}'/2 \rfloor \quad (24)$$

And then, the original pixel pair could be recovered according to Eqs. (7)-(8).

*D. Difference Expansion in Image Shares*

DE-IS ensures that the shareholder could embed additional data into his own image share and extract it directly from the marked share. Then DE-IS marked share could be losslessly recovered. As for the $i$-th shareholder, $i \in \{1, ..., n\}$, DE-IS could be implemented on the image share $S_i$ or the marked share $S_i'$. In this section, we take $S_i'$ and $K_{SIS}^i$ as an example to elaborate processes of DE-IS.

*1) Overflow Constraints*

As for the ciphertext $c_{x,y}^{i\prime} \in S_i'$ and $ID_{x,y}^i \in K_{SIS}^i$, ($x=1,..., 512$; $y=1,..., 512$), we aim to embed one bit additional data $b_s$ into the pair ($c_{x,y}^{i\prime}$, $ID_{x,y}^i$) by difference expansion. To avoid overflow, the constraints are deduced first. According to Eq. (18):

$$0 \leqslant c_{x,y}^{i\prime} < ID_{x,y}^i \quad (25)$$

The difference, denoted as $h_L$, can be computed:

$$h_L = ID_{x,y}^i - c_{x,y}^{i\prime} \quad (26)$$

Then the constraint of $h_L$ is obtained:

$$0 \leqslant 2 \times h_L < q_i \quad (27)$$

$LSB(ID_{x,y}^i)$ would be set 0 if the constraint in Eq. (27) was not satisfied to label the available ciphertext for DE-IS.

*2) DE-IS Marked Image Share Generation*

Step 1: Randomly encrypt the additional data $b_s$ by using a random $k$ from the data hiding key $k$ to obtain the to-be-embedded data $b_L$:

$$b_L = k \oplus b_s \quad (28)$$

Step 2: If $LSB(ID_{x,y}^i)=1$, ($c_{x,y}^i{'}$, $ID_{x,y}^i$) should be an available pair for DE. The difference $h_L$ is computed by Eq. (26). DE-IS marked ciphertext is obtained:

$$c_{x,y}^i{''} = 2 \times h_L + b_L \quad (29)$$

Repeat above Step 1-2 onto all the available ciphertext. DE-IS marked share $S_i''$ could be obtained, where $c_{x,y}^i{''} \in S_i'$, $ID_{x,y}^i \in K_{SIS}^i$, $x=1,\ldots,512$; $y=1,\ldots,512$.

*3) Data Extraction*

With DE-IS marked share $S_i''$, additional data could be directly extracted from it.

Step 1: If $LSB(ID_{x,y}^i)=0$, there is no data embedded; If $LSB(ID_{x,y}^i)=1$, $b_L = LSB(c_{x,y}^i{'})$.

Step 2: Recover the additional data $b_s$ by using $k$:

$$b_s = k \oplus b_L \quad (30)$$

*4) Image Share Recovery*

With DE-IS marked share $S_i''$ and the SIS key $K_{SIS}^i$, the pair ($c_{x,y}^i{'}$, $ID_{x,y}^i$) can be recovered from ($c_{x,y}^i{''}$, $ID_{x,y}^i$):

If $LSB(ID_{x,y}^i)=0$, $c_{x,y}^i{'} = c_{x,y}^i{''}$, $ID_{x,y}^i = ID_{x,y}^i + 1$; or if $LSB(ID_{x,y}^i)=1$, $c_{x,y}^i{'} = ID_{x,y}^i - \lfloor c_{x,y}^i{''}/2 \rfloor$.

## V. Theoretical Analysis and Experimental Results

### A. Correctness

The correctness of the proposed scheme includes the lossless recovery of the image shares and the accurate extraction of the additional data. The test images are 512×512 8-bit grayscale images from image libraries, USC-SIPI (*http://sipi.usc.edu/database/database.php?volume=misc*). All the experiments were all implemented on MATLAB2010b with a 64-bit single core CPU (i7-6800K) @ 3.40GHz. Experimental results of six test images (Fig. 5) were demonstrated in this section.

*Parameters setting*: $n=7$, $t=5$; $q_0=257$; $q=\{457, 461, 463, 467, 479, 487, 491, 499, 503, 509\}$; $0 \le h_{fid} \le 10$. Each shareholder was distributed a SIS key, where the primes were all from $q$ and satisfied the conditions in Eqs. (1)-(3).

Experimental results at different stages of the proposed scheme on image Lena under the above settings are shown in Fig. 6: Secret image Lena was shared into $n$ shares, one of which is shown in Fig. 6(a); then the image share in Fig. 6(a) was embedded by HDE-ED, and HDE-ED marked share was shown in Fig. 6(b); HDE-ED marked share was secondly embedded by DE-IS to obtain DE-IS marked share as shown in Fig. 6(c); The recovered share from DE-IS marked share is shown in Fig. 6(d), which is the same as Fig. 6(b); The directly reconstructed image from HDE-ED marked shares is shown in Fig. 6(e); The recovered image is shown in Fig. 6(f).

*1) Image Recovery*

In HDE-ED, the number of the available pixel pairs determines the EC of HDE-ED in an image. The available pixel pairs are obtained according to the constraints in Eqs. (11)-(12). To avoid the overflow caused by homomorphic operations among ciphertext, we performed homomorphic plus one on the ciphertext of the available pairs in each image share. If there is no overflow, $M_{ava}$ remains unchanged; otherwise, if overflow occurs, the label "1" is replaced by "0" in $M_{ava}$. The maximum EC of HDE-ED of the six test images are listed in Table II denoted as EC$_1$. After HDE-ED embedding, the dealer could obtain the marked image $I'$ by directly reconstructing no less than $t$ ($t=5$) marked shares.

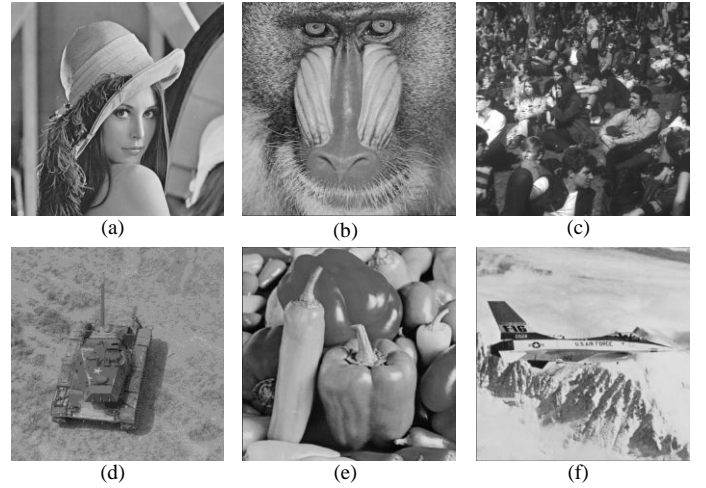

Fig. 5. Test images. (a) Lena; (b) Baboon; (c) Crowd; (d) Tank; (e) Peppers; (f) Plane.

TABLE II
THE PSNR$_1$ – PSNR$_3$ (*dB*) AND EC$_1$-EC$_2$ (*bit*) AT $h_{fid}$=10

| Image   | EC$_1$  | PSNR$_1$ | PSNR$_2$ | EC$_2$   | PSNR$_3$ |
|---------|---------|----------|----------|----------|----------|
| Lena    | 110165  | 42.3171  | ∞        | 805386   | ∞        |
| Baboon  | 69256   | 41.9894  | ∞        | 1068993  | ∞        |
| Crowd   | 103882  | 42.6764  | ∞        | 911363   | ∞        |
| Tank    | 108963  | 41.4472  | ∞        | 941326   | ∞        |
| Peppers | 110548  | 41.5025  | ∞        | 892745   | ∞        |
| Plane   | 114834  | 43.2519  | ∞        | 777365   | ∞        |

We calculated the Peak Signal to Noise Ratio (PSNR) between $I'$ under EC$_1$ and $I$, which are also recorded in Table II as PSNR$_1$. With the marked image, the dealer could operate DE recovery to obtain the restored image. The PSNRs of the restored images were calculated and listed in Table II as PSNR$_2$. It should be noted that there would be no image but noise obtained when using any less than $t$ shares for reconstruction, which is determined by the threshold of secret sharing.

In DE-IS, the EC of an image share is determined by the number of the available ciphertext meeting the constraint in Eq. (27). Due to the random generation of the ciphertext in shares, there is much uncertainty in the EC of DE-IS for different image shares. To demonstrate the performance of DE-IS, we enumerate the average EC in Table II denoted as EC$_2$. EC$_2$ is obtained by repeating the operations of image sharing and DE-IS embedding 10 times per test image and calculating the average sum of the capacities of the $n$ ($n=7$) shares from one image.

DE-IS ensures that the shareholder could recover the marked share into the same one before embedding. We enumerate the average PSNR between the recovered share and the share before DE-IS embedding in Table II, which are denoted as PSNR$_3$.

In Table II, the PSNR$_2$ and PSNR$_3$ are all "∞", which means that the secret images can be losslessly recovered after HDE-ED and the image shares can also be losslessly recovered after DE-IS. Besides, PSNR$_1$ indicates that there are distortions in the directly reconstructed images after HDE-ED.

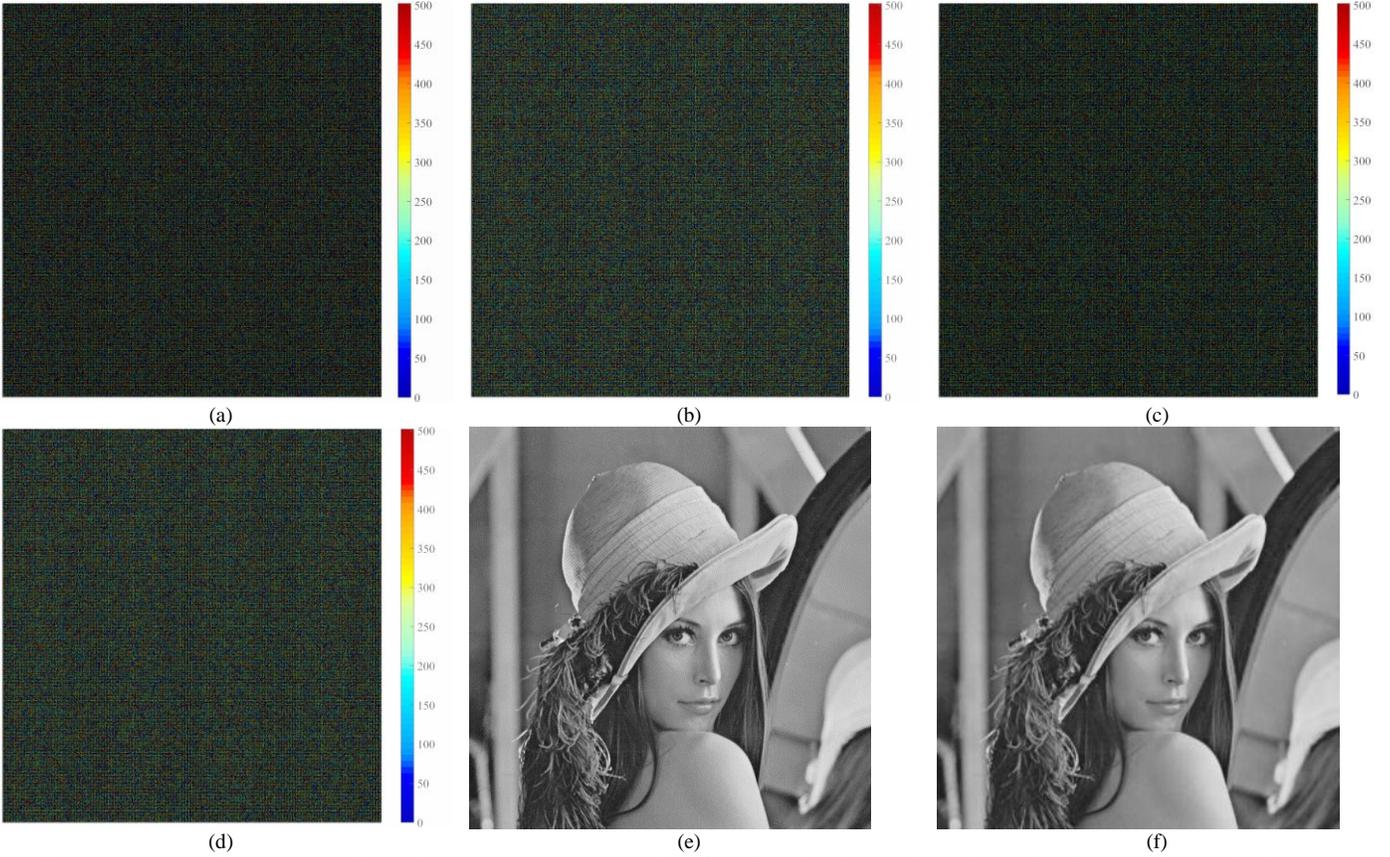

Fig. 6. Experimental results on Lena: (a) One of the $n$ image shares of Lena; (b) HDE-ED marked share with $h_{fid}=10$; (c) DE-IS marked shares; (d) The recovered share from DE-IS marked share; (e) The directly reconstructed Lena from HDE-ED marked shares; (f) The recovered Lena.

TABLE III
PSNR1 (dB) VERSUS EC$_1$(bits) AT DIFFERENT $h_{fid}$

| Image | $h_{fid}=5$ | | $h_{fid}=3$ | | $h_{fid}=2$ | | $h_{fid}=1$ | | $h_{fid}=0$ | |
|---|---|---|---|---|---|---|---|---|---|---|
| | EC$_1$ | PSNR1 | EC$_1$ | PSNR1 | EC$_1$ | PSNR1 | EC$_1$ | PSNR1 | EC$_1$ | PSNR1 |
| Lena | 86605 | 45.6073 | 65303 | 49.3394 | 50232 | 52.2932 | 32104 | 56.6637 | 11434 | 64.7369 |
| Baboon | 42522 | 47.9412 | 28553 | 52.5559 | 20702 | 55.9498 | 12464 | 60.7332 | 4210 | 69.0304 |
| Crowd | 86962 | 47.0505 | 73240 | 50.1791 | 64164 | 52.1256 | 46810 | 55.9208 | 26504 | 61.0445 |
| Plane | 100746 | 46.1582 | 85200 | 48.9055 | 71114 | 51.2680 | 50764 | 54.8316 | 19966 | 62.3590 |
| Peppers | 80017 | 45.4989 | 56523 | 49.7460 | 42091 | 52.9404 | 26044 | 57.5117 | 8791 | 65.9745 |
| Tank | 77887 | 45.6201 | 53520 | 50.3489 | 43832 | 52.5652 | 20988 | 60.6655 | 16843 | 63.1004 |

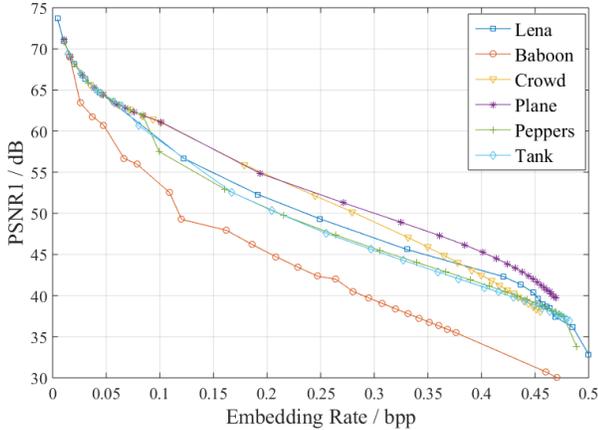

Fig. 7. The relationship between PSNR$_1$ and embedding rates by HDE-ED

We next analyze the PSNR$_1$ of the marked images at different EC by HDE-ED. As shown in Table III, we list the EC$_1$ and PSNRs of the six test images at different $h_{fid}$. Fig. 7 shows the relationship between PSNR$_1$ and the embedding rates (ER) by HDE-ED. ER is the amount of additional data embedded per pixel (bpp). Then a comparison of the performances of PSNR$_1$ at different ER was made among six RDH-ED methods [9][21][22][32][33][36] and the proposed. As shown in Fig. 8, we demonstrated the comparison results from test images Lena (Fig. 8a) and Plane (Fig. 8b). The results show that there are fewer distortions in the marked images by HDE-ED than the existing RDH-ED schemes.

*2) Data Extraction*

HDE-ED ensures that data could be extracted from the marked image after reconstruction, while DE-IS ensures that the data could be extracted from any a marked share before reconstruction. Fig. 9(a)-(b) shows the errors by comparing the extracted data by HDE-ED and DE-IS with the to-be-embedded data bit by bit with an EC of $10^5$ bits. The results show that the extraction accuracy is 100%. In order to achieve the separability of RDH-ED, same additional data could be embedded twice by HDE-ED and DE-IS successively. Due to the more than one shares existing, there would be a much higher EC of DE-IS than that of HED-ED according to the results in Table II.

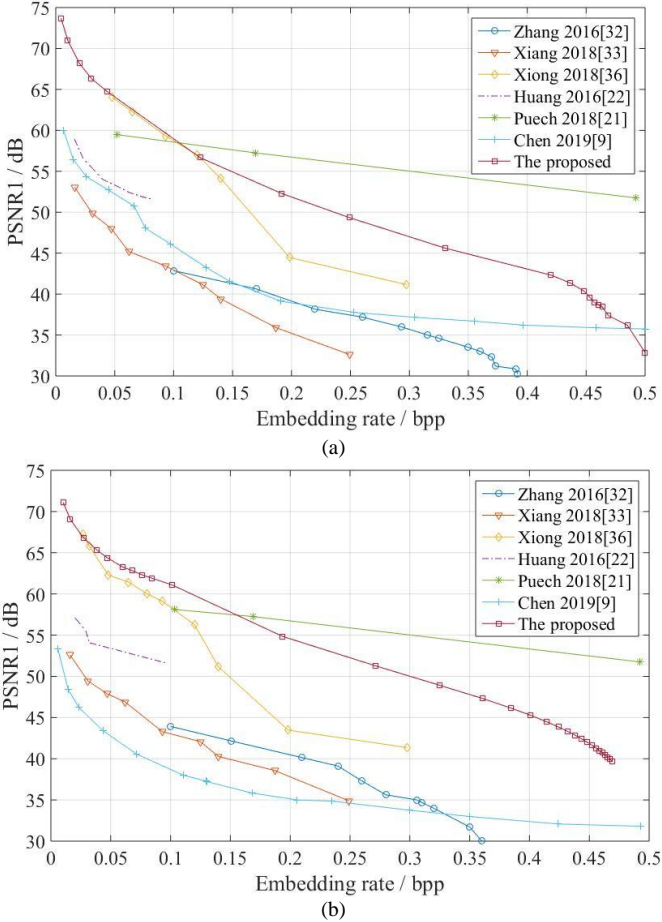

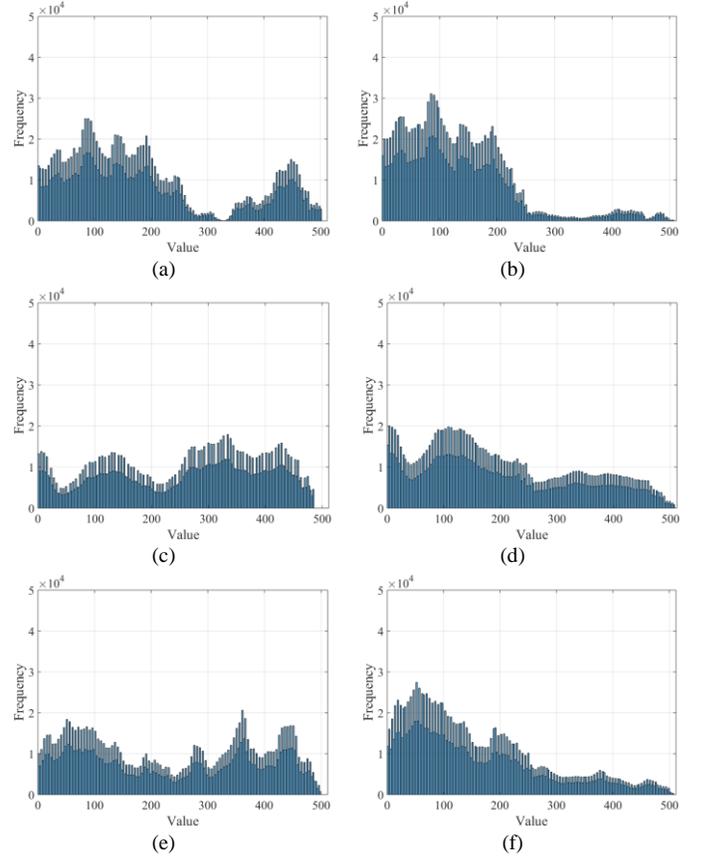

Fig. 8. PSNR$_1$ (dB) of HDE-DE at different ER (bpp) on (a) Lena; (b) Plane.

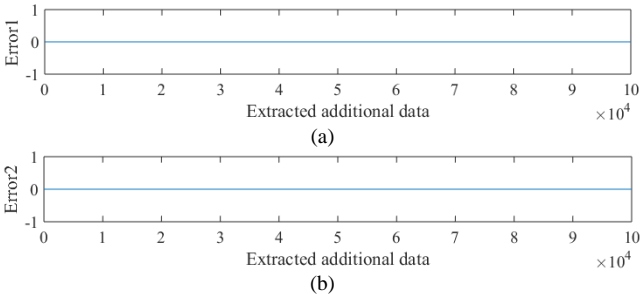

Fig. 9. Errors of the extracted data: (a) Error$_1$ from HDE-ED extraction on the marked images; (b) Error$_2$ from DE extraction on the marked shares.

*B. Security*

Security mainly includes two aspects [1][34][35]: *a)* Data hiding should not weaken the encryption intensity or leave any risks of security cracking. *b)* Any information about the additional data cannot be obtained without the data-hiding key.

*1) HDE-ED*

For HDE-ED, the embedding processes are based on the homomorphism of CRT secret sharing. All the homomorphic operations are on the ciphertext of the image shares and would not reveal any information about the pixels or SIS keys. After HDE-ED, the marked shares are still legal image shares, which would not weaken the encryption intensity of SIS system. Therefore, HDE-ED could ensure the security of RDH-ED.

Fig. 10. Histograms of ciphertext in the image shares before and after DE-IS: (a) from Lena before DE-IS; (b) from Baboon after DE-IS; (c) from Baboon before DE-IS; (d) from Baboon after DE-IS; (e) from Peppers before DE-IS; (f) from Peppers after DE-IS.

*2) DE-IS*

For DE-IS, the additional data is encrypted by data-hiding key before embedding, thus ensuring the secrecy of additional data. However, during data hiding processes, the ciphertext of the image share would get modified. To verify the security of DE-IS, it is necessary to analyze the statistical characteristics of the ciphertext before and after DE-IS embedding. It is necessary to remain the statistical characteristics of the images shares stable before and after embedding [21], [34], [35].

To demonstrate the distribution characteristics intuitively, the histograms of ciphertext of different image shares before and after DE-IS embedding were shown in Fig. 9, in which we mainly enumerate the histograms of Lena, Baboon, and Peppers with the max EC. Due to the uneven distribution the modulus in $q$ follow, the ciphertext in the image shares is not fully randomly and uniformly distributed as shown in Fig. 10. However, randomness still exists in the process of generating ciphertext in SIS, based on which the pixel content is kept secured against any attackers without the SIS keys [37] (More relevant security analysis of CRT secret sharing in [37], [38]).

To further analyze the impact of DE-IS on the distribution of ciphertext quantitatively, the *correlation* and the *information*

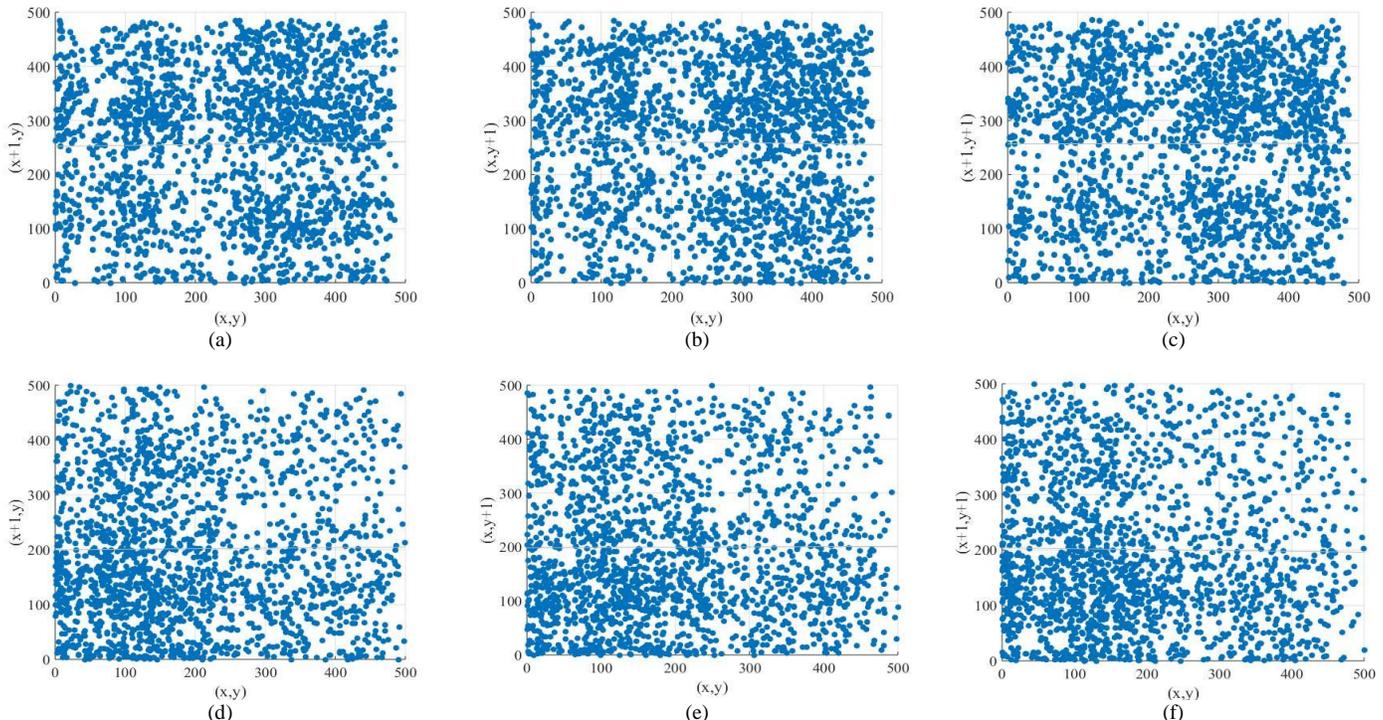

Fig. 11. Correlation analysis on Baboon in three direction before and after DE-IS embedding: (a) Horizontally adjacent before DE-IS; (b) Vertically adjacent before DE-IS; (c) Diagonally adjacent before DE-IS; (d) Horizontally adjacent after DE-IS; (e) Vertically adjacent after DE-IS; (f) Diagonally adjacent after DE-IS.

TABLE IV
ADJACENT CORRELATION OF IMAGE SHARES BEFORE AND AFTER DE-IS ON TEST IMAGES

| Image | Vertically adjacent correlation | | Horizontally adjacent correlation | | Diagonally adjacent correlation | |
|---|---|---|---|---|---|---|
| | Before | After | Before | After | Before | After |
| Lena | -0.00864 | 0.00734 | 0.01444 | -0.02030 | 0.02062 | -0.03574 |
| Baboon | -0.01772 | -0.00632 | 0.01746 | 0.01706 | 0.00273 | 0.01611 |
| Crowd | 0.00504 | -0.01368 | -0.00549 | 0.00107 | 0.02591 | 0.01190 |
| Plane | -0.02706 | 0.01879 | 0.03778 | -0.01074 | 0.00677 | -0.01043 |
| Peppers | 0.04002 | 0.00350 | -0.03823 | 0.03030 | 0.00518 | 0.01792 |
| Tank | 0.01741 | -0.02913 | -0.00735 | -0.01513 | -0.00765 | -0.01948 |

TABLE V
AVERAGE INFORMATION ENTROPY OF IMAGE SHARES BEFORE AND AFTER DE-IS ON TEST IMAGES

| Image | $S_1$ | $S_1''$ | $S_2$ | $S_2''$ | $S_3$ | $S_3''$ | $S_4$ | $S_4''$ | $S_5$ | $S_5''$ | $S_6$ | $S_6''$ | $S_7$ | $S_7''$ |
|---|---|---|---|---|---|---|---|---|---|---|---|---|---|---|
| Lena | 6.7711 | 7.7211 | 6.7759 | 7.7228 | 6.7687 | 7.7239 | 6.7815 | 7.7229 | 6.7759 | 7.7213 | 6.7773 | 7.7233 | 6.7755 | 7.7230 |
| Baboon | 4.4101 | 6.3438 | 4.4192 | 6.3458 | 4.4294 | 6.3503 | 4.4287 | 6.3399 | 4.4090 | 6.3378 | 4.4413 | 6.3510 | 4.4230 | 6.3501 |
| Crowd | 5.8298 | 7.6314 | 5.8266 | 7.6291 | 5.8266 | 7.6284 | 5.8188 | 7.6306 | 5.8287 | 7.6326 | 5.8265 | 7.6304 | 5.8247 | 7.6282 |
| Tank | 4.7318 | 7.3704 | 4.3645 | 7.3722 | 4.3679 | 7.3688 | 4.3485 | 7.3700 | 4.3672 | 7.3733 | 4.3727 | 7.3677 | 4.3656 | 7.3679 |
| Peppers | 5.1049 | 7.1284 | 5.0969 | 7.1285 | 5.0721 | 7.1178 | 5.0860 | 7.1304 | 5.0885 | 7.1147 | 5.1122 | 7.1316 | 5.0950 | 7.1336 |
| Plane | 5.1991 | 7.4597 | 5.4870 | 7.4538 | 5.4748 | 7.4480 | 5.4856 | 7.4545 | 5.4889 | 7.4486 | 5.5028 | 7.4614 | 5.4885 | 7.4582 |

*entropy* of the ciphertext are calculated below.

The *correlation* between adjacent pixels is an important indicator for analyzing the effect of encryption, by which the correlation of adjacent pixels is reduced as much as possible. The adjacent correlation is divided into horizontal, vertical, and diagonal correlation. The correlation coefficient between variables $u$ and $v$ is denoted as $r_{u,v} \in$ [-1, 1]. Let $N_s$ be the amount of randomly sampling data for computing $r_{u,v}$:

$$E(u) = \frac{1}{N_s} \sum_{i=1}^{N_s} u_i \quad (31)$$

$$D(u) = \frac{1}{N_s} \sum_{i=1}^{N_s} (u_i - E(u))^2 \quad (32)$$

$$Cov(u,v) = \frac{1}{N_s} \sum_{i=1}^{N_s} (u_i - E(u))(v_i - E(v)) \quad (33)$$

$$r_{u,v} = \frac{Cov(u,v)}{\sqrt{D(u)} \times \sqrt{D(v)}} \quad (34)$$

Ideally, zero correlation should be achieved. In practice, it could be considered that there is a significant correlation between $u$ and $v$ when $|r_{u,v}| > 0.6$. Usually, the absolute value of the correlation between adjacent pixels in a natural image can reach more than 0.6. Image encryption is required to reduce the absolute value of the coefficient of the adjacent ciphertext to be below 0.2 [40].

We calculated the correlation coefficients with 2000 samples of adjacent ciphertext ($N_s$=2000) from the image shares and the marked shares by DE-IS. The results were recorded in Table V and the absolute values of the correlation coefficients are all kept below 0.0400.

To visually reflect the statistical results of the correlation, Fig. 11 shows the adjacent correlations of the three directions sampled from the shares of Baboon before and after DE-IS. The stronger the correlation is, the more dots appear near the diagonal $x=y$ of the coordinate axis. It can be seen from the distribution of dots in Fig. 11 that the marked shares by DE-IS maintains the random distribution.

Encryption is required to increase the *average information entropy* of encrypted data as high as possible. We calculated the average information entropy of the original image shares and DE-IS marked shares respectively.

In the ciphertext set with a value range of [0, $q$-1], ciphertext could be regarded as signal $a_i$ ($i$=0, 1, ..., $q$-1). The probability of $a_i$ is represented by the statistical frequency of $a_i$, denoted as $h(a_i)$. The average information entropy of $a_i$ is denoted as $H$:

$$H = -\sum_{i=0}^{q} h(a_i) \log h(a_i) \quad (35)$$

Table V lists the average information entropy calculated from all the shares before and after DE-IS embedding. It can be seen that the average information entropy of the marked ciphertext tends to be higher than the original shares, thus ensuring the security of the marked ciphertext.

### C. Efficiency

#### 1) Storage Consumption

In the multiparty system of SIS, though the total amount of ciphertext increases linearly with the number of participating shareholders, which is determined by the function of the SIS system, *i.e.*, *to realize access control and share the risk of decryption failure of the system*, the storage consumption of a single shareholder remains unchanged. Therefore, we mainly analyze the storage consumption of a single shareholder in this section to discuss the efficiency first. The total storage consumption of a multiparty system is $n$ times the consumption of a shareholder ($n \geq 2$).

The storage consumption of a shareholder is mainly related to the ciphertext expansion of the scheme. In [35], blowup factor (BF) of ciphertext expansion was introduced. If the blowup factor is denoted as $l_{BF}$, it means that 1 bit plaintext will be encrypted into ciphertext of $l_{BF}$ bits.

RDH-ED based on public-key cryptosystem has ciphertext extension while symmetric encryption based RDH-ED has no ciphertext extension, which is determined by the encryption principle rather than the data hiding methods. As discussed in [35] and [1], BF of stream encryption based RDH-ED [3]-[6], [19]-[22] is 1 while BF of RDH-ED based on Paillier encryption [27]-[33] and LWE encryption[1], [34]-[35] is larger than $O(10^2)$.

Due to the constraints of certain parameters for secret sharing, there would be pixels that cannot be recovered when the parameters for sharing are set less than 255. On the other hand, it will cause certain ciphertext extension when the parameters are set greater than 255, which is a common problem of image secret sharing schemes.

Schemes in [7], [9] selected the parameters less than 255 to ensure that no ciphertext extension introduced into the shares.

TABLE VI
COMPARISON OF COMPLEXITY

| Encryption | Typical schemes | Complexity |
|---|---|---|
| Stream encryption | [3]- [6], [19]- [22] | $O(N)$ |
| Paillier encryption | [27]-[33] | $O(N^3)$ |
| LWE encryption | [34]- [35] | $O(N^2)$ |
| FHE encryption | [1] | $O(N^2)$- $O(N^3)$ |
| Secret sharing | [7], [9], the proposed | $O(N)$ |

BF of the image sharing schemes in [7], [9] is 1. However in [7], [9], the overflowed pixels must be labeled before image sharing and the remainder of the overflowed pixels would be transmitted as side information for image recovery.

In this paper, considering the technical demand of efficiency and lossless recovery, the modulus in $q$ is set within [$2^w$, $2^{w+1}$], $w$ =8. Therefore, BF of the proposed scheme is $l_{BF}$ =1.125.

In practice, it is also possible to reduce all the ciphertext extension of the proposed scheme by setting $w$=7. Before image sharing, the MSBs of pixels are first losslessly compressed and then embedded into image shares by DE-IS. Then BF of the ciphertext in each shareholder would be reduced to 1. The detail implementation would be our future work.

#### 2) Computational complexity

Let $N$ be the length of plaintext, the computational complexities of different RDH-ED schemes are compared in Table VI.

#### 3) Embedding Capacity

In HDE-ED, the embedding rate is related to the setting of $h_{fid}$ and the content of test images. When $h_{fid}$ is fixed, the smoother the image is, the higher ER can be obtained. When $h_{fid}$ is set "∞", ER of test images can reach close to 0.5000bpp according the experiments in Section V.A.

In DE-IS, the total embedding capacity is related to the available ciphertext for embedding of all the $n$ image shares. According to the results in Table II, we can obtain the average embedding capacity of each share:

$$EC_{DE-IS} = \frac{EC_2}{n} \quad (36)$$

Considering the blowup factor of encryption, the average embedding rate of ciphertext of DE-IS can be calculated:

$$ER_{DE-IS} = \frac{EC_{DE-IS}}{512 \times 512 \times 9} \quad (37)$$

$ER_{DE-IS}$ is 0.0545bpb according to the results in Table II, namely, 0.0545 bits additional data could be embedded into per bit of ciphertext in average.

## VI. CONCLUSION

In this paper, a separable RDH-ED scheme based on CRT is presented aimed at providing technical supports for secret image sharing application. S*eparability* of RDH-ED is discussed based on the application that RDH-ED provides for secret image sharing. To realize separability, two data hiding methods are proposed: one is *homomorphic difference expansion in encrypted domain* that supports data extraction from the reconstructed image by utilizing the addition homomorphism of CRT secret sharing; the other is *difference expansion in image shares* that supports the data extraction directly from the marked shares without image reconstruction.

The performances of the proposed scheme in correctness,

security, and efficiency are analyzed theoretically and experimentally. Experimental results demonstrate that the proposed scheme maintains (*t*, *n*) threshold of CRT secret sharing and obtains a better embedding capacity compared with existing RDH-ED algorithms. Future investigation will focus on introducing more complicated secret image sharing schemes and optimizing the efficiency of RDH-ED schemes.

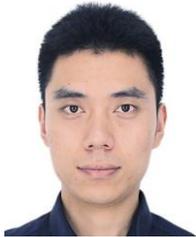
**Yan Ke** received the M.S. degree in cryptography from Engineering University of PAP, Xi'an, China in 2016 and the Ph.D. degree in cryptography at Engineering University of PAP. Currently, he has been with the Counterterrorism Command & Information Engineering Joint Lab in Urumqi Campus of Engineering University of PAP. He is also with the Key Laboratory of Network and Information Security under Chinese People Armed Police Force in the School of Cryptography Engineering in Engineering University of PAP. His research interest includes information hiding, lattice based cryptography, multi-party computing security.

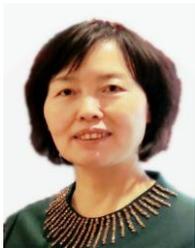
**Minqing Zhang** received the M.S. degree in computer science & application from Northwestern Polytechnical University, Xi'an, China, in 2001, and Ph.D. degree in network & information security from Northwestern Polytechnical University, Xi'an, China, in 2016. Currently, she has been with the Key Laboratory of Network and Information Security under Chinese People Armed Police Force in the School of Cryptography Engineering in Engineering University of PAP as a professor. Her research interests include cryptography, and trusted computation.

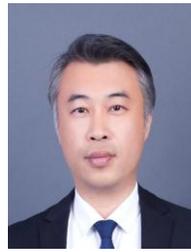
**Xinpeng Zhang** received the B.S. degree in computational mathematics from Jilin University, China, in 1995, and the M.E. and Ph.D. degrees in communication and information system from Shanghai University, China, in 2001 and 2004, respectively. Since 2004, he was with the faculty of the School of Communication and Information Engineering, Shanghai University, where he is currently a Professor. He is also with the faculty of the School of Computer Science, Fudan University. He was with The State University of New York at Binghamton as a Visiting Scholar from 2010 to 2011, and also with Konstanz University as an experienced Researcher, sponsored by the Alexander von Humboldt Foundation from 2011 to 2012. His research interests include multimedia security, image processing, and digital forensics. He has published over 200 papers in these areas. He was an Associate Editor of the IEEE Transactions on Information Forensics and Security (T-FIS) from 2014 to 2017.

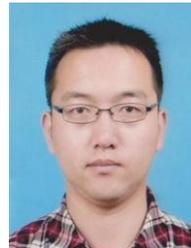
**Jia Liu** received the M.S. degree in cryptography from Engineering University of PAP, Xi'an, China, in 2007 and Ph.D. degree in neural network and machine learning from Shanghai Jiao Tong University, Shanghai, China, in 2012. Currently, he has been with the Key Laboratory of Network and Information Security under PAP in the School of Cryptography Engineering in Engineering University of PAP as an associate professor. His research interests include pattern recognition and image processing.

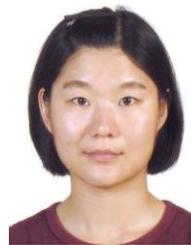
**Tingting Su** received the B.S. degree in information research & security from Engineering University of PAP, Xi'an, China in 2010 and the M.S. degree in cryptography from Engineering University of PAP, Xi'an, China in 2013. Currently, she has been with the Key Laboratory of Network and Information Security under PAP in the School of Cryptography Engineering in Engineering University of PAP. Her research interest includes mathematics statistics and cryptography.

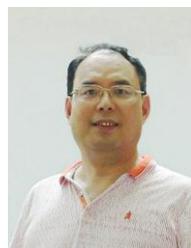
**Xiaoyuan Yang** received the B.S. degree in applied mathematics from Xidian University, Xi'an, China, in 1982 and the M.S. degree in cryptography & encoding theory from Xidian University, Xi'an, China, in 1991. Currently, he has been with the Key Laboratory of Network and Information Security under PAP in the School of Cryptography Engineering in Engineering University of PAP as a professor. His research interests include cryptography and trusted computation.